\documentclass[9pt, twocolumn]{article}
\usepackage{times,bm}
\usepackage[affil-it,]{authblk}
\usepackage[pagebackref=false,colorlinks=true,linkcolor=acsblue,citecolor=red,urlcolor=acsblue]{hyperref}
\usepackage{geometry}
\usepackage[switch*, displaymath,pagewise, mathlines]{lineno}
\usepackage{graphicx}
\usepackage{cite}
\usepackage{amsmath}
\usepackage{amsfonts}
\usepackage{amssymb}
\usepackage{slashed}
\usepackage{subfloat}
\usepackage{slashed}
\usepackage{color}
\usepackage{float}
\usepackage{multirow,multicol}
\usepackage{tabulary}
\usepackage[flushleft]{threeparttable}
\usepackage{pgfplots,subfigure}
\usepackage{xcolor}
\numberwithin{equation}{section}
\usepackage{tikz}
\usepackage{ulem}
\usepackage{hyperref}
\usepackage{nameref}
\usepackage{comment}

\newlength\wttlinewidth
\usepgflibrary{arrows}
\usetikzlibrary{shapes.callouts}
\tikzset{
	level/.style   = { thick, },
	connect/.style = { dotted, red   },
	notice/.style  = { draw, rectangle callout, callout relative pointer={#1} },
	label/.style   = { text width=2cm }
}

\definecolor{acsblue}{RGB}{17,76,139}
\definecolor{shadecolor}{RGB}{255,241,204}
\usepackage{letltxmacro}
\makeatletter
\let\oldr@@t\r@@t
\def\r@@t#1#2{%
	\setbox0=\hbox{$\oldr@@t#1{#2\,}$}\dimen0=\ht0
	\advance\dimen0-0.2\ht0
	\setbox2=\hbox{\vrule height\ht0 depth -\dimen0}%
	{\box0\lower0.4pt\box2}}
\LetLtxMacro{\oldsqrt}{\sqrt}
\renewcommand*{\sqrt}[2][\ ]{\oldsqrt[#1]{#2}}
\makeatother
\usepackage{environ}

\begin{document}

\newcommand{{\ri}}{{\rm{i}}}
\newcommand{{\Psibar}}{{\bar{\Psi}}}
\newcommand*\var{\mathit}

\title{\mdseries{Investigating quantum criticality through charged scalar fields near the BTZ black hole horizon}}
\author{ \textit {Abdullah Guvendi}$^{\ 1}$\footnote{\textit{E-mail: abdullah.guvendi@erzurum.edu.tr}}~,~ \textit {Omar Mustafa}$^{\ 2}$\footnote{\textit{E-mail: omar.mustafa@emu.edu.tr (Corr. Author)} }  \\
	\small \textit {$^{\ 1}$ Department of Basic Sciences, Erzurum Technical University, 25050, Erzurum,
		Turkiye}\\
	\small \textit {$^{\ 2}$ Department of Physics, Eastern Mediterranean University, 99628, G. Magusa, north Cyprus, Mersin 10 - Turkiye}\\}
\date{}

\maketitle

\begin{abstract}

We examine a charged scalar field with a position-dependent mass \( m(\rho) = m_0 + \mathcal{S}(\rho) \), where \(\mathcal{S}(\rho)\) represents a Lorentz scalar potential, near a BTZ black hole in the presence of an external magnetic field. By deriving the Klein-Gordon equation for this setup, we explore two scenarios: (i) a mass-modified scalar field with \(m(\rho) = m_0 - a/\rho\) (an exactly solvable case), and (ii) a scenario involving both mass modification and an external magnetic field (conditionally exactly solvable). We identify quantum critical points (QCPs) associated with the coupling constant \(a\). In the first scenario, for massless charged scalar fields, critical points occur at \(a = n + 1/2\) for all radial quantum numbers \(n \geq 0\) and magnetic quantum numbers \(|m| \geq 0\). In the second scenario, these critical points shift to \(a = n + 3/2\) for \(n \geq 0\) and \(|m| > 0\), with the case \(m = 0\) excluded. For massive scalar fields, QCPs emerge at \(a = (n + 1/2)/2\), leading to non-propagating fields at zero frequency. At these QCPs, the field frequencies drop to zero, marking transitions from stable oscillatory modes to non-propagating states. Below the critical points, the system exhibits instability, characterized by negative imaginary frequencies that suggest rapid decay and high dissipation. Above the critical points, the modes stabilize and propagate, indicating a transition to a superconducting-like phase, where dissipation vanishes and stable excitations dominate.

\end{abstract}

\begin{small}
\begin{center}
\textit{Keywords: BTZ black hole, Quantum critical points, AdS/CFT correspondence, Holographic superconductivity, Magnetic field}	
\end{center}
\end{small}

\section{\mdseries{Introduction}}\label{sec1}

Black holes provide a unique framework for exploring the connections between quantum mechanics, gravity, and thermodynamics, particularly in the regions surrounding their event horizons. Their distinct structures provide a natural environment for testing fundamental theories that extend beyond classical general relativity. Within the context of the Kerr/Conformal Field Theory correspondence, black hole horizons serve as crucial connections between gravitational physics and conformal symmetry, a relationship extensively explored in previous studies \cite{encodes}. Among the various solutions to Einstein's field equations, the Banados-Teitelboim-Zanelli (BTZ) black hole \cite{BTZ} emerges as a particularly powerful model. This \(2+1\)-dimensional black hole, characterized by a negative cosmological constant (\(\Lambda = -1/\ell^2\)), shares essential properties with its higher-dimensional counterparts—such as mass, angular momentum, and charge—while offering a more tractable framework for studying quantum phenomena. Despite its lower dimensionality, the BTZ black hole effectively captures key aspects of black hole thermodynamics and quantum field behavior, making it an invaluable tool for probing quantum effects near horizons \cite{carlip, corichi}.

The AdS/CFT (anti-de Sitter/Conformal Field Theory) correspondence has revolutionized our understanding of gravitational theories, particularly those formulated in AdS spaces, and their relationship to quantum field theories on the boundary \cite{hol1, hol2, hol3}. This holographic duality has facilitated the development of models exploring phase transitions and critical phenomena, especially in strongly interacting systems. A notable application lies in the study of holographic superconductors, which describe the coupling between black holes in AdS space and charged quantum fields. These models provide profound insights into superconductivity in strongly correlated materials. Below a critical temperature or energy, charged scalar fields interacting with black holes may undergo spontaneous symmetry breaking, leading to the emergence of a superconducting phase, described by real oscillation modes \cite{hol4, hol5, hol6, hol7, hol8, hol9, hol10, hol11}. This framework enhances our understanding of phase transitions, quantum criticality, and the nature of strong coupling in superconducting materials \cite{hol01,hol02,hol03,hol04}.

Quantum critical points (QCPs) signify transitions between distinct phases of matter at zero temperature or energy. These transitions are vital for explaining a diverse array of exotic phenomena, such as non-Fermi liquid behavior and other unconventional phases that deviate from classical theories of matter \cite{inkof,davison,sachdev}. The behavior of scalar fields near the event horizon of BTZ black holes offers a "well-defined" setting for studying quantum criticality. Specifically, analyzing charged scalar fields subjected to external influences, such as magnetic fields, yields valuable insights into the quantum critical behavior that can emerge in these strongly coupled systems. These insights are especially significant in scenarios where standard perturbative methods fail, which is often the case in systems with strong interactions or near criticality. To date, no exact solutions have been reported that analyze quantum criticality for the relativistic quantum fields governing charged scalar fields in the presence of external fields near black hole horizons. Therefore, deriving analytical results for charged scalar fields in the near-horizon region of a BTZ black hole is essential for understanding quantum critical phenomena in these environments.

The Klein-Gordon equation serves as a foundational tool for studying scalar quantum fields in curved spacetime. Solutions to this equation within black hole backgrounds are crucial for understanding phenomena such as particle emission rates, spectra, and quantum radiation \cite{reff}. Specifically, they enable the investigation of Hawking radiation and its implications for the relationship between quantum mechanics, thermodynamics, and general relativity \cite{H1, K1}. Exploring scalar fields in the near-horizon region of black holes like the BTZ black hole provides valuable insights into the behavior of quantum fields in strong gravitational fields. These findings have broader implications not only for black hole physics but also for related fields such as gravitational wave astronomy, where scalar field perturbations can model signals from black hole mergers. Furthermore, they offer pathways to explore profound questions regarding spacetime structure, information loss, and high-energy processes near black hole horizons \cite{K1}. When analyzing quantum fields, interactions with external forces can be modeled through three primary mechanisms: electromagnetic minimal coupling \cite{Greiner}, non-minimal coupling \cite{NH1,NH2,NH3} and the introduction of a Lorentz scalar potential \cite{AO1,AO2,AO}. In the minimal coupling framework, the charged scalar field interacts with an external electromagnetic field through the vector potential \(\mathcal{A}_\mu = (\phi, \mathbf{A})\), where \(\phi\) represents the scalar component and \(\mathbf{A}\) represents the vector component. This interaction modifies the Klein-Gordon equation by replacing the ordinary derivative \(\partial_\mu\) with the covariant derivative \(D_\mu = \partial_\mu + ie\mathcal{A}_\mu\), ensuring gauge invariance and providing a quantum electrodynamics-consistent description of the interaction between charged particles and external fields \cite{Greiner}. Conversely, the Lorentz scalar potential modifies the mass term in the Klein-Gordon equation by introducing a position-dependent scalar potential \(\mathcal{S}(\rho)\), leading to a modified mass term of the form \(m_{0} \rightarrow m_{0} + \mathcal{S}(\rho)\) \cite{AO1,AO2}. This adjustment impacts the local energy of the particle without altering its gauge properties \cite{AO}. These two complementary approaches facilitate the study of scalar fields interacting with external fields, each offering unique insights into the underlying physics. Notably, the position-dependent mass function may be derived from various physical contexts. In the bulk theory, this mass function can be interpreted as a result of dynamical mass generation due to interactions, such as the influence of external fields or Lorentz scalar potentials. Within the holographic context, the position-dependent mass signifies the effective coupling that modifies the dynamics of scalar fields near the horizon of a black hole. Such modifications can lead to rich physical phenomena, including mass generation, instability, and critical behavior, which are pivotal for understanding phase transitions and the emergence of condensates in the dual field theory. Notably, this kind of mass modification may result in a normal phase characterized by high dissipation and instability. Conversely, for values above the critical points, the modes transition to real oscillatory states, indicating a return to stability and the presence of propagating excitations, corresponding to a superconducting phase. This behavior can be observed through the analytical solution of Lorentz-invariant wave equations, reflecting fundamental changes in the excitation spectrum and indicating that the system undergoes a quantum phase transition. The emergence of zero-frequency modes at the QCPs corresponds to critical phenomena associated with phase transitions, highlighting the relationship between stability, decay, and the physical properties of the system, particularly within the context of holographic superconductivity.

This paper focuses on examining the behavior of a charged scalar field in the near-horizon region of a BTZ black hole, subject to an external magnetic field and a position-dependent mass. The primary objective is to investigate the emergence of quantum critical points, which manifest as zero-frequency states in this system. To achieve this, we solve the generalized Klein-Gordon equation with a position-dependent mass and systematically analyze the conditions leading to these critical points. In Section \ref{sec2}, we introduce the governing equations and derive the relevant second-order wave equation. Section \ref{sec2.1} presents an exact solution for the wave equation in the absence of an external magnetic field, modifying the mass term as \(m_0 \rightarrow m_0 - \frac{a}{\rho}\), where \(a\) represents the coupling constant \cite{AO}. Section \ref{sec2.2} extends the analysis to include the effects of an external magnetic field, yielding conditionally exact solutions. We conclude by summarizing the quantum critical behavior observed in these cases and discussing the broader implications for quantum field dynamics near black hole horizons. Throughout this paper, we adopt natural units, where \(G = 1 = \hbar = c\).

\section{\mdseries{Wave equation and its solutions}} \label{sec2}

In this section, we provide an overview of the BTZ black hole's near-horizon background and derive a wave equation for a charged scalar field with a position-dependent mass and an external magnetic field. The Euclidean solution for the BTZ black hole with positive mass \( M \) and zero angular momentum is given by \cite{corichi}:
\begin{equation}
ds^2 = \frac{r^2 - M\ell^2}{\ell^2} d\tilde{\tau}^2 + \frac{\ell^2}{r^2 - M\ell^2} dr^2 + r^2 d\theta^2. \label{eq1}
\end{equation}
Substituting \(\ell \rightarrow i\ell\) and \(\tilde{\tau} \rightarrow i\tilde{\tau}\), and transforming \(r \rightarrow T\) and \(\tilde{\tau} \rightarrow T\), we obtain:
\begin{equation}
ds^2 = -\frac{\ell^2}{T^2 + M\ell^2} dT^2 + \frac{T^2 + M\ell^2}{\ell^2} dR^2 + T^2 d\theta^2. \label{eq2}
\end{equation}
This spacetime structure resembles a cosmological "big bang" with a singularity at \( T = 0 \) and is the dual of the BTZ black hole solution. Considering negative mass \( M = -\alpha^2 \) within \((0,1]\), the solution, known as a "particle" solution, shows conical singularities at \( r = 0 \) with a deficit angle \(\Omega = 2\pi(1 - \alpha) \). When \(\alpha = 1\), the solution becomes a singularity-free global \( AdS_3 \) spacetime. Using established identifications, we derive the line element:
\begin{equation}
ds^2 = -\frac{\alpha^2\ell^2 - r^2}{\ell^2} d\tilde{\tau}^2 + \frac{\ell^2}{\alpha^2\ell^2 - r^2} dr^2 + r^2 d\theta^2. \label{eq3}
\end{equation}
This describes spacetime with a cosmological horizon at \( r = r_c = \alpha\ell \) and a conical singularity at \( r = 0 \) with angle \(\Omega\). By introducing the coordinate transformation \( \rho^2 = \ell^2 - \frac{r^2}{\alpha^2} \) and substituting \(\tilde{\tau} \rightarrow t\) and \(\theta \rightarrow \phi\), we obtain the metric that describes the near-horizon region of the BTZ black hole \cite{corichi}. The resulting metric is \cite{corichi}:
\begin{equation}
ds^2 \approx -\frac{\alpha^2 \rho^2}{\ell^2} \, dt^2 + d\rho^2 + \alpha^2 \ell^2 \, d\phi^2. \label{eq4}
\end{equation}
Accordingly, the covariant metric tensor \( g_{\mu\nu} \) and its inverse \( g^{\mu\nu} \) can be written
\begin{equation*}
\begin{split}
&g_{\mu\nu} = \text{diag}\left(-\frac{\alpha^2 \rho^2}{\ell^2}, 1, \alpha^2 \ell^2\right),\\
&g^{\mu\nu} = \text{diag}\left(-\frac{\ell^2}{\alpha^2 \rho^2}, 1, \frac{1}{\alpha^2 \ell^2}\right).
\end{split}
\end{equation*}
For a charged scalar field with a position-dependent mass and an external magnetic field, where \(\mathcal{A}_{\phi} = -\frac{\mathcal{B}_{0}}{2}\rho\) (a magnetic field that satisfies the inhomogeneous Maxwell equation, as discussed in \cite{AO1, AO2, AO}), the corresponding Klein-Gordon equation near the horizon of the BTZ black hole becomes \cite{NH1}:
\begin{equation}
\begin{split}
&\frac{1}{\sqrt{-g}}\mathcal{D}_{\mu}\left[\sqrt{-g} g^{\mu\nu} \mathcal{D}_{\nu} \Psi\right]=m_0(\rho)^2\,\Psi\left(x^{\mu}\right),\\
&\mathcal{D}_{\mu} = \partial_{\mu} - ie\mathcal{A}_{\mu}, \quad \mathcal{D}_{\nu} = \partial_{\nu} - ie\mathcal{A}_{\nu}, \label{eq5}
\end{split}
\end{equation}
where \(g = \det(g_{\mu\nu})\), \(e\) is the electric charge of the scalar field, and \(m(\rho) = m_0 + \mathcal{S}(\rho)\), with \(m_0\) being the scalar boson's rest mass and \(\mathcal{S}(\rho)\) the Lorentz scalar potential. Decomposing the scalar field as \(\Psi(x^{\mu}) = \exp(-i\omega t) \exp(im\phi) \psi(\rho)\), where \(\omega\) is the relativistic frequency and \(m\) is the magnetic quantum number, we use \(A_{\mu} = (0, 0, A_{\phi})\) with \(A_{\phi} = -\frac{\mathcal{B}_{0}}{2}\rho\). The resulting wave equation is:
\begin{equation}
\begin{split}
&\psi'' + \frac{1}{\rho}\psi' + \left[\frac{\Omega^2}{\rho^2} - \left(\bar{m} + \mathcal{B}\rho\right)^2 - m_0(\rho)^2\right]\psi = 0, \\
&\Omega = \frac{\omega \ell}{\alpha}, \quad \bar{m} = \frac{m}{\alpha\ell}, \quad \mathcal{B} = \frac{e \mathcal{B}_{0}}{2\alpha\ell}. \label{eq6}
\end{split}
\end{equation}
In the following sections, we will seek analytical solutions to this wave equation under different scenarios.

\subsection{\mdseries{Exact results for the case where \(m_0(\rho)\rightarrow m_{0}-a/\rho\) and \(\mathcal{B}_{0}=0\)}}\label{sec2.1}

In this section, we present an exact solution to Eq. (\ref{eq6}) by modifying the mass, which changes depending on position, $m_0(\rho)  \rightarrow m_{0}-a/\rho$, where $a$ denotes the coupling constant \cite{AO}. In this particular case, Eq. (\ref{eq6}) is written explicitly as the following
\begin{equation}
\psi^{''}+\frac{1}{\rho}\psi^{'}+\left[\frac{\Omega^2}{\rho^2}-\bar{m}^2-(m_{0}-\frac{a}{\rho})^2\right]\psi=0.\label{eq6.1}      
\end{equation}
By considering a new change of variable, $z=2\sqrt{m^2_{0}+\bar{m}^2}\,\rho$ \footnote{Note that $z\rightsquigarrow 0 (\infty)$ as $\rho\rightsquigarrow 0 (\infty)$.}, and then by employing an ansatz, $\psi(z)=\frac{\tilde{\psi}(z)}{\sqrt{z}}$, one obtains the following equation
\begin{equation}
\left[\partial_{z}^2+\frac{\tilde{\mu}}{z}+\frac{\frac{1}{4}-\tilde{\nu}^2}{z^2}-\frac{1}{4} \right]\tilde{\psi}(z)=0. \label{eq7}
\end{equation}
Here, the parameters are defined as:
\begin{equation*}
\tilde{\mu}=\frac{m_{0}\,a}{\sqrt{\bar{m}^2+m^2_{0}}},\quad \tilde{\nu}=\sqrt{a^2-\Omega^2}.
\end{equation*}
 The solution to Eq. (\ref{eq7}) can be expressed using the Confluent Hypergeometric function, given by
\begin{equation*}\tilde{\psi}(z)=e^{-\frac{z}{2}}\,z^{\frac{1}{2}+\tilde{\nu}}\  _1F_{1}(\frac{1}{2}+\tilde{\nu}-\tilde{\mu}, 1+2\tilde{\nu}; z).
\end{equation*}
Here, it is worth underlining that our focus is on a system confined within the near-horizon region, where the regular solution must exhibit exponential decay as the spatial coordinate extends infinitely, specifically $\tilde{\psi}(z)\propto e^{-z}$. Nonetheless, we acknowledge the importance of comprehending the asymptotic behavior of the function:
\begin{equation*}
\begin{split}
&_1F_{1}(\frac{1}{2}+\tilde{\nu}-\tilde{\mu}, 1+2\tilde{\nu}; z)\\
&\approx  \frac{\Gamma(1+2\tilde{\nu})}{\Gamma(\frac{1}{2}+\tilde{\nu}-\tilde{\mu})}\frac{e^{z}}{z^{(\frac{1}{2}+\tilde{\nu}+\tilde{\mu})}}\left[1+\mathcal{O}(|z|^{-1}) \right], 
\end{split}
\end{equation*}
which diverges as $z\rightsquigarrow\infty$ \cite{NH4}. In our pursuit of uncovering the regular solution (around the $z=0$) to the wave equation, our objective is to establish it based on the condition: $\frac{1}{2}+\tilde{\nu}-\tilde{\mu}=-n$ \cite{NH4}, where $n$ represents the radial quantum number $(n=0,1,2...)$. This condition forms the foundation for deriving the relativistic frequency expression relevant to the system under investigation
\begin{equation}
\omega_{n m}=-i\frac{\alpha}{\ell}\sqrt{a^2\left[\frac{m^2_0}{(m^2_0+\bar{m}^2)}-1\right]+\bar{n}^2-\frac{2am_0\bar{n}}{\sqrt{m^2_0+\bar{m}^2}}},\label{eq8}
\end{equation}
in which $\bar{n}=n+\frac{1}{2}$. Notably, two distinct and interesting sets of solutions emerge from equation (\ref{eq8}). The first set corresponds to \(S\)-states (i.e., \(\bar{m} = 0\)), yielding
\begin{equation}
\omega_{n,I} = -i \frac{\alpha}{\ell} \sqrt{\bar{n}^2 - 2a\bar{n}}. \label{eq9}
\end{equation}

\begin{figure}[ht]
\centering
\includegraphics[scale=0.60]{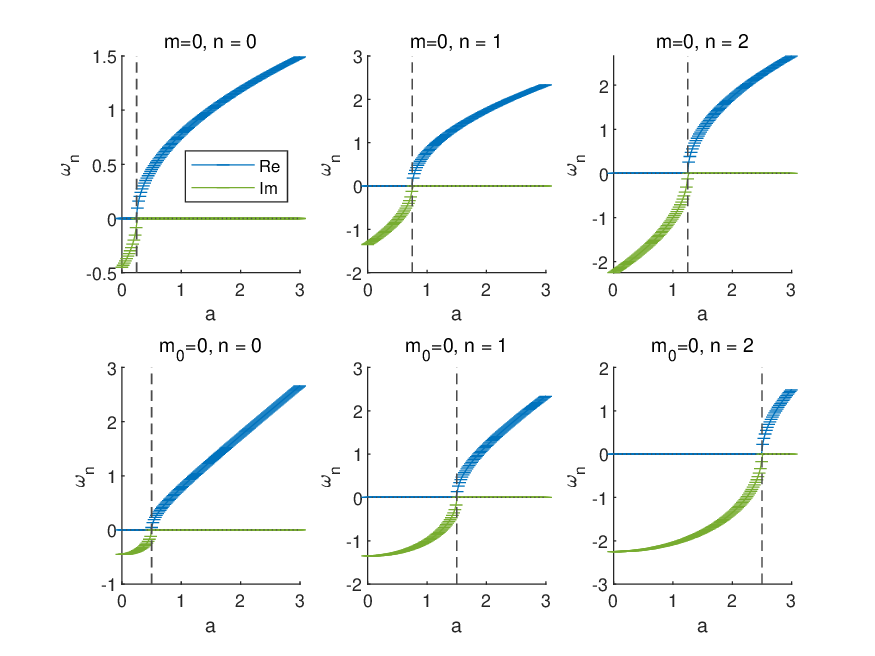}\\
\caption{This figure illustrates the real (blue line) and imaginary parts (green line) of the frequency \(\omega\) as functions of the coupling constant \(a\) for two sets of quantum states. The first row represents the S-states (\(m = 0\)), while the second row corresponds to massless scalar fields (\(m_0 = 0\)). Each subplot displays the frequencies for quantum numbers \(n = 0, 1, 2\). The quantum critical points are indicated by dashed vertical lines, marking the values of \(a\) where the frequencies become zero, signifying phase transitions and stability changes in the system. Here $\alpha=0.9$ and $\ell=1$.}
\label{fig:a}
\end{figure} 

The second set applies to massless scalar fields with \(m_0 = 0\), valid for all radial quantum numbers \(n \geq 0\) and magnetic quantum numbers \(|m| \geq 0\), and is given by
\begin{equation}
\omega_{n,II} = -i \frac{\alpha}{\ell} \sqrt{\bar{n}^2 - a^2}. \label{eq9.1}
\end{equation}
For both sets, we identify quantum critical points (QCPs) where \(\omega_{n,I} = 0\) at \(a = a_{c,I} = \bar{n}/2\) and \(\omega_{n,II} = 0\) at \(a = a_{c,II} = \bar{n}\). At these critical values, denoted as \(a_{c,I}\) and \(a_{c,II}\) (see also Figures \ref{fig:a}), the states cease to evolve with time and effectively disappear (refer to Figure \ref{fig:c}). For values of \(a < a_{c,I}\) and \(a < a_{c,II}\), the corresponding states become unstable and decay over time. For example, \(\omega_{0,I} = 0\) when \(a = \frac{1}{4}\), indicating that \(a = 0.25\) is a QCP. This suggests that the system cannot sustain itself over time if \(a = 0.25\) (since \(\Psi \propto \exp(-i \omega t)\)). For \(a < \frac{1}{4}\), the \(n=0\) quantum state is unstable and decays over time, with the decay time given by \(\tau_{n} = \frac{1}{|\Im \omega_{n}|} \Rightarrow \tau_{0} = \frac{\ell}{\alpha \sqrt{1/4 - a}}\), depending on the coupling constant \(a\) and the spacetime parameters \(\alpha\) and \(\ell\). Conversely, for \(a > \frac{1}{4}\), a real oscillation mode becomes viable, characterized by the frequency \(\omega_{0} = \frac{\alpha}{\ell} \sqrt{a^2 - \frac{1}{4}}\). 

Here, it is evident that the thermalization of the system is dominated by the ground state. Thus, in the limit where $a \to 0$, the imaginary part of the frequency, $\Im \omega_{0}$, enables us to recover the formal Hawking temperature ($T_H$), as detailed in \cite{corichi}. Specifically:
\begin{equation}
\begin{split}
&T_H \propto \frac{|\Im \omega_{0}|}{\pi} \Rightarrow T_H = \frac{\alpha}{2\pi \ell} = \frac{\sqrt{M}}{2\pi \ell}\\
&\text{or}\\
&T_H = \frac{r_c}{2\pi \ell^2},
\end{split}
\end{equation}
where \(M = \alpha^2 \in (0,1)\) denotes the black hole mass, and $r_c$ is the cosmological horizon with $r_c = \alpha \ell$ \cite{corichi}. The identification of QCPs is essential for understanding the evolution of a system, as these critical points mark the transition from stable oscillatory modes to non-propagating states, signaling a substantial shift in its physical properties. For values below the critical threshold, the corresponding quantum states become unstable, leading to decay over time, which can be interpreted as a loss of coherence similar to that observed in superconducting materials when they transition to a non-superconducting phase. Conversely, above the critical points, the emergence of real oscillation modes suggests the formation of stable excitations akin to coherent condensates in superconductors, emphasizing the connection between gravitational phenomena and strongly correlated quantum systems.

\subsection{\mdseries{Conditionally exact results for \(m_0(\rho)\rightarrow m_0-a/\rho\) and \(\mathcal{B}_{0}\neq0\)}}\label{sec2.2}

\begin{figure}[ht]
    \centering
    \includegraphics[scale=0.60]{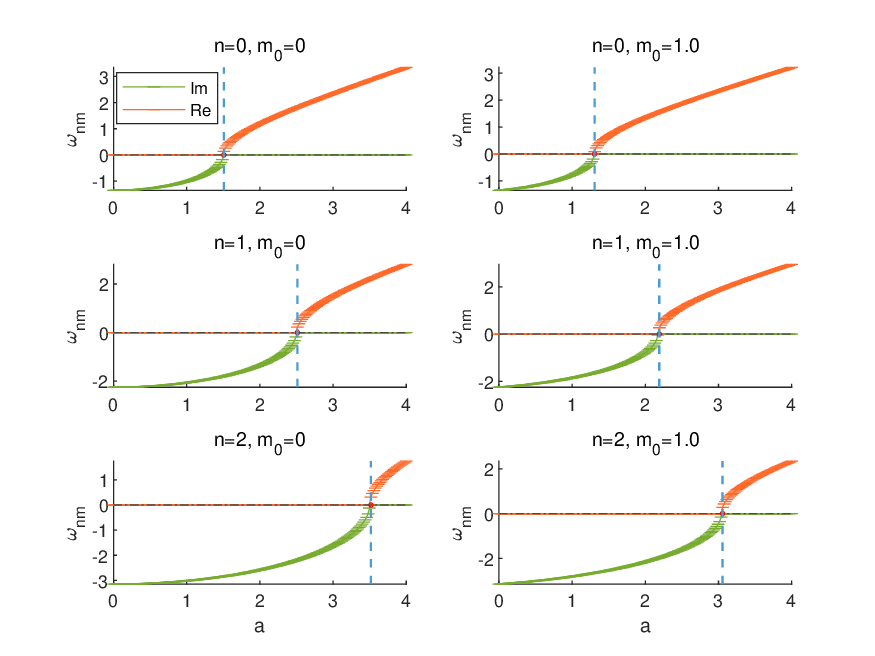}\\
    \caption{Real and imaginary frequency behavior for different quantum states \( n \) and mass parameters \( m_0 \). The green curves indicate the imaginary (Im) component of the frequency \( \omega \), and the orange curves depict the real (Re) component. The quantum critical points are identified by vertical dashed blue lines, and these points mark the transition between decaying and oscillatory modes. The horizontal dotted line represents the zero-frequency level. Here, $\alpha=0.9$, $\ell=1$ and $m=1$.}
    \label{fig:b}
\end{figure}

\begin{figure}[ht]
    \centering
    \includegraphics[scale=0.55]{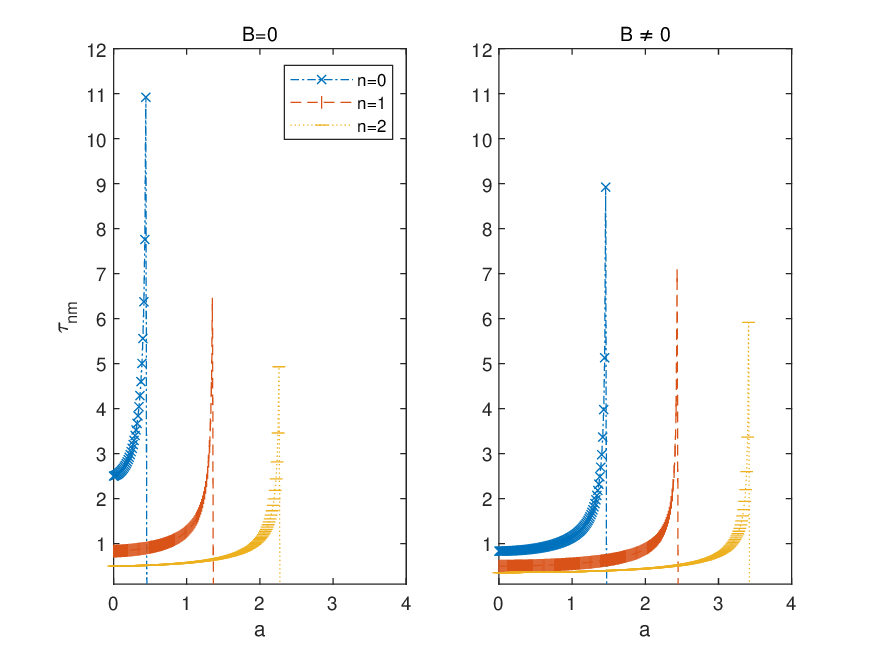}\\
    \caption{The decay time $\tau_{nm}$ for varying $a$ values. Here, $m_0=0.1$, $m=1$, $\alpha=0.9$ and $\ell=1$}
    \label{fig:c}
\end{figure}

In this section, we unveil a solution that precisely addresses Eq. (\ref{eq6}) when we incorporate a modification wherein the mass adjusts with position, $m_0(\rho)\rightarrow m_0-a/\rho$, aligning with the attractive Coulomb potential within the external magnetic field's influence. Under this circumstance, the equation transforms into:
\begin{equation}
\psi^{''}+\frac{1}{\rho}\psi^{'}+\left[\frac{\Omega^2}{\rho^2}-\left(\bar{m}+\mathcal{B}\rho\right)^2-(m_0-\frac{a}{\rho})^2\right]\psi=0,\label{eq9.1}      
\end{equation}
and this equation possesses a solution in the form of biconfluent Heun functions/series and is given by
\begin{equation}
\begin{split}
&\Psi\left(\rho\right)=\exp(-\frac{\mathcal{B}\rho^2}{2}-|\bar{m}|\rho)\,\mathcal{H}_{B}\left(\tilde{\alpha},\tilde{\beta},\tilde{\gamma},\tilde{\delta},\sqrt{\mathcal{B}}\rho\right),\\
&\tilde{\alpha}=2\sqrt{a^2-\Omega^2},\quad \tilde{\beta}=\frac{2|\bar{m}|}{\sqrt{\mathcal{B}}},\quad \tilde{\gamma}=-\frac{m_0^2}{\mathcal{B}},\\
& \tilde{\delta}=-\frac{4m_0}{\sqrt{\mathcal{B}}}.\label{eq10}
\end{split}
\end{equation}
The biconfluent Heun function, denoted as $\mathcal{H}_{B}(\tilde{\alpha},\tilde{\beta},\tilde{\gamma},\tilde{\delta}; \sqrt{\mathcal{B}}\rho)$, requires truncation to a polynomial of order $n+1\geq 1$ concerning the variable $y$, where $y=\sqrt{\mathcal{B}}\rho$. This condition is detailed in a new approach introduced in the Appendix of Mustafa in \cite{omar-1}, and subsequently employed by Mustafa and colleagues \cite{omar-2,omar-3}. Essentially, this approach necessitates limiting the biconfluent Heun functions/series to a polynomial of order $n+1\geq 1$ under the fulfillment of two conditions:
\begin{equation}
\begin{split}
&\tilde{\gamma}=2(n+1)+\tilde{\alpha} \Rightarrow \\
&\frac{m_0^2}{\mathcal{B}}= -2(n+1) -2\sqrt{a^2-\Omega^2},\label{eq11}
\end{split}
\end{equation} 
and
\begin{equation}
\begin{split}
&\tilde{\delta}=-\tilde{\beta}(2n+\tilde{\alpha}+3)\Rightarrow \\ & a\,m_0 =|\bar{m}|(2n+2\sqrt{a^2-\Omega^2}+3).\label{eq12} 
\end{split}
\end{equation}
Accordingly, one arrives at the following expression 
\begin{equation}
\mathcal{B}=\frac{m_0^2 |\bar{m}|}{|\bar{m}|-a\,m_0} \Rightarrow \frac{e\mathcal{B}_{0}}{2\alpha\ell}=\frac{m_0^2 |\bar{m}|}{|\bar{m}|-a\,m_0}. \label{om1} 
\end{equation}
This result would suggest that neither  $\mathcal{B}_0$ nor the magnetic quantum number $m$ are allowed to be zeros. This is a common price one has to pay when conditional exact solvability is involved in the process. Nevertheless, such conditionally exact solvability allows us to elaborate on a vast number of $m\neq 0$-states with $n\geq 0$  and provide some reliable analysis on the spectroscopic structure. One should also observe that the result in (\ref{om1}) provides a parametric correlation that facilitates conditional exact solvability of the problem at hand. As a result, provided $m\neq 0$, we can deduce the relativistic frequency expression for a charged scalar field in the vicinity of the BTZ black hole's horizon
\begin{equation}
\begin{split}
\omega_{nm }= -i\frac{\alpha}{\ell}\sqrt{n^2_{\dagger}+a^2\left[\frac{m_0^2}{4\bar{m}^2}-1\right]-\frac{m_0\, a}{2|\bar{m}|}n_{\dagger}}.\label{eq13}
\end{split}
\end{equation}
When $m_0=0$ (i.e., massless particles), this expression simplifies to:
\begin{equation}
\omega_{n}= -i\frac{\alpha}{\ell}\sqrt{n^2_{\dagger}-a^2},\quad n_{\dagger}=n+\frac{3}{2}.\label{eq14}
\end{equation}
In Eq. (\ref{eq14}), it is evident that all values of $ a=a_c= n_\dagger$ form QCPs for which $ a_c<  n_\dagger$  the corresponding states become unstable and decay over time. For example, for $n=0$ we clearly observe that $\omega_{0}=0$ when $a=\frac{3}{2}$ (QCP). For $a<\frac{3}{2}$, this state cannot be steady and decays over time with a decay time $\tau_{0}=\frac{1}{|\Im \omega_{0}|}$, which simplifies to $\tau_{0}=\frac{\ell}{\alpha\sqrt{(3/2)^2-a^2}}$. This decay time depends on the coupling strength constant $a$ and the spacetime parameters $\alpha$ and $\ell$. Conversely, when $a>\frac{3}{2}$, this mode becomes a real oscillation mode with a frequency given by $\omega_{0}=\frac{\alpha}{\ell}\sqrt{a^2-(3/2)^2}$ (see Eq. (\ref{eq14})). Our result, given by Eq. (\ref{eq13}), shows that the evolution of the corresponding system explicitly depends on the value of the parameter \(a\), in addition to other factors such as the rest mass of the particle, spacetime parameters, and quantum numbers. Furthermore, we observe that the QCPs shift according to the rest mass of the scalar particle, as illustrated in Figure \ref{fig:b}. In both scenarios, a closer look at the \(m_0=0\) cases reveals that the QCPs (\(a_{c}\)) determined for the first scenario shift to \(a_{c}+1\) when the effects of the external magnetic field are considered. In other words, the "effective" ground state (\(\bar{n}=1/2\)) identified in the first scenario shifts to (\(n_\dagger=3/2\)) when the magnetic field effects are taken into account. These effects are clearly illustrated in Figures \ref{fig:a}, \ref{fig:b}, and \ref{fig:c}. Figure \ref{fig:c} also shows that the decay time of the damped modes can be relatively longer as the system approaches QCPs, given \(a < a_c\) in each scenario. However, the system's frequency drops to zero upon reaching the quantum critical points (QCPs), indicating that the system disappears over time at these points. Figure \ref{fig:c} also shows that the thermalization of the system is dominated by the ground state in each scenario.
As the system approaches its ground state, the frequency expressions derived from each scenario can be related to the modified Hawking temperature \(T^{'}_H\). In the second scenario, which accounts for the effects of the magnetic field, we can derive the modified Hawking temperature for \(n=0\) and \(m_0=0\) as \(T^{'}_H \propto \frac{|\Im \omega_{0}|}{\pi} = \frac{\alpha}{\pi \ell} \sqrt{\frac{9}{4} - a^2}\). Notably, this modified temperature approaches zero as the critical value \(a\) is reached, specifically at \(a_{c} = \frac{3}{2}\). This result indicates that the temperature tends to zero when the system reaches the QCPs identified in each scenario, leading to the conclusion that the black hole ceases to emit thermal radiation. Such a phenomenon implies a transition to a zero-temperature phase, characteristic of a quantum phase transition, where the system is described by purely quantum states dominated by quantum effects. It is known that the Hawking temperature is fundamentally linked to the thermal radiation emitted by black holes due to quantum fluctuations occurring at the event horizon, which contributes to their evaporation. In the context of holographic superconductivity, this temperature is pivotal through the AdS/CFT correspondence, where the temperature of the black hole's horizon corresponds to that of the boundary theory. As the system cools towards the Hawking temperature, it undergoes a phase transition from a normal phase to a superconducting state, paralleling the behavior observed in conventional superconductors. This transition is driven by the gravitational effects in the bulk spacetime, emphasizing the connections between gravity, thermodynamics, and condensed matter physics in the context of holography.

\section{\mdseries{Summary and discussions}} \label{sec3}

In this manuscript, we investigated the evolution of a charged scalar particle with a position-dependent mass in the presence of an external magnetic field near the horizon of the BTZ black hole. We employed the Klein-Gordon equation to derive a general second-order wave equation with a position-dependent mass and sought analytical solutions under two scenarios: (i) with mass modification \(m_{0} \rightarrow m_0(\rho)= m_0 - a/\rho\), introducing an attractive Coulomb-like Lorentz scalar interaction (as discussed in \cite{AO}) without an external magnetic field, and (ii) including both mass modification \(m_{0} \rightarrow m_0 - a/\rho\) and the effects of the external magnetic field. 

In the first scenario, we obtained an exact solution to the wave equation in terms of Confluent Hypergeometric functions, which allowed us to derive the relativistic frequency (\(\omega_{nm}\)) as given by Eq. (\ref{eq8}). This enabled us to identify QCPs for two cases: For \(S\)-states (i.e., \(\bar{m}=0\)), the frequency is \(\omega_{n,I}=-i\frac{\alpha}{\ell}\sqrt{\bar{n}^2-2a\bar{n}}\), with QCPs at \(a=a_{c,I}=\bar{n}/2\) where \(\omega_{n,I}=0\). For massless scalar fields (\(m_0=0\)) with all radial \(n\geq0\) and magnetic \(|m|\geq0\) quantum numbers, the frequency is \(\omega_{n,II}=-i\frac{\alpha}{\ell}\sqrt{\bar{n}^2-a^2}\), with critical points at \(a=a_{c,II}=\bar{n}\). At these quantum critical values, \(a_{c,I}\) and \(a_{c,II}\), the states cease to evolve over time and effectively disappear; for \(a<a_{c,I}\) and \(a<a_{c,II}\), the states become unstable and decay over time. For example, in the \(S\)-state, \(\omega_{0,I}=0\) when \(a=a_{c,I}=\frac{1}{4}\), indicating a QCP. This suggests that for \(a=0.25\), the system cannot persist (since \(\Psi \propto \exp(-i\omega t)\)). For \(a<0.25\), the state is unstable with a decay time \(\tau_{0}=\frac{\ell}{\alpha\sqrt{1/4-a}}\) (see also Fig. \ref{fig:c}), dependent on the coupling constant \(a\) and the spacetime parameters \(\alpha\) and \(\ell\). Conversely, for \(a>0.25\), a real oscillation mode is feasible with a frequency \(\omega_{0,I}=\frac{\alpha}{\ell}\sqrt{a-0.25}\) (see also Fig. \ref{fig:a}).

In the second scenario, we derived a conditionally exact solution \cite{AO,omar-1,omar-2,omar-3} using biconfluent Heun functions. For \(m \neq 0\), the relativistic frequency expression for a charged scalar field with a position-dependent mass in the presence of the magnetic field near the BTZ black hole's horizon is:
\[
\omega_{nm}= -i\frac{\alpha}{\ell}\sqrt{n^2_{\dagger}+a^2\left[\frac{m_0^2}{4\bar{m}^2}-1\right]-\frac{m_0\, a}{2\bar{m}}n_{\dagger}},
\]
which simplifies to 
\[
\omega_{n}= -i\frac{\alpha}{\ell}\sqrt{n^2_{\dagger}-a^2},\quad n_{\dagger}=n+\frac{3}{2},
\]
when \(m_0=0\). Here, \(\omega_{0}=0\) when \(a=\frac{3}{2}\), marking a QCP. For \(a<\frac{3}{2}\), the state cannot remain stable and decays over time with a decay time \(\tau_{0}=\frac{\ell}{\alpha\sqrt{9/4-a^2}}\) (see also Fig. \ref{fig:c}), dependent on the coupling constant \(a\) and spacetime parameters \(\alpha\) and \(\ell\). For \(a>\frac{3}{2}\), the mode becomes a real oscillation mode with frequency \(\omega_{0}=\frac{\alpha}{\ell}\sqrt{a^2-9/4}\) (see also Fig. \ref{fig:b}). Figure \ref{fig:c} shows that the decay time of damped modes approaches a maximum value as the system reaches nearly QCPs where the frequency (\(\omega\)) and hence energy are zero, indicating zero-temperature critical behavior. As the system's thermalization is dominated by its ground state, the corresponding frequency expressions obtained for the each analyzed scenarios can be associated with the modified Hawking temperature. Accordingly, in all cases, the modified Hawking temperature can be formulated. Significantly, the modified temperature approaches zero as the critical value \(a\) is attained, notably at \(a=a_{c}\). This observation indicates that the temperature tends to vanish when the system reaches the QCPs identified in each scenario, suggesting that the black hole stops emit thermal radiation. This marks a transition to a zero-temperature phase typical of a quantum phase transition, where quantum states dominate the system, heavily influenced by quantum effects. In the realm of holographic superconductivity, the temperature holds critical importance through the AdS/CFT correspondence, linking the temperature at the black hole's horizon to that of the boundary theory. As the system cools, it transitions from a normal state to a superconducting phase, mirroring the behaviors exhibited by conventional superconductors. In the context of holographic superconductivity and conductivity, our results may have significant implications. A real oscillation mode corresponds to a stable excitation in the boundary theory, often associated with a normal phase in holographic superconductivity. For conductivity, such modes imply well-defined excitations with non-zero energy, which are crucial for understanding the system's physical response. The appearance of zero-frequency modes indicates QCPs, associated with critical behavior. This is particularly relevant for holographic superconductivity, marking transitions between different phases or critical points where quantum phase transitions occur. When the frequency mode becomes purely imaginary (negative in our case), it corresponds to a damped mode, indicating instability or a phase where the system is no longer superconducting or exhibits high dissipation. For conductivity, such modes suggest the absence of stable low-energy excitations, essential for understanding superconductivity or its breakdown. Overall, real oscillation modes are most directly useful for describing stable, propagating excitations that contribute to the system's physical conductivity, while zero-energy and damped modes provide valuable insights into critical phenomena and system stability.

On the other hand, graphene-like structures can be characterized by relativistic-like equations because of their linear dispersion relation. By solving these equations, we can explore the effective dielectric properties of such materials and understand how particles interact with external electromagnetic fields. The dielectric function $\epsilon(\omega)$ of these materials is derived from their response to an external electromagnetic field, which involves calculating particle polarization and using it to determine permittivity. For instance, to investigate the effective dielectric properties of a wormhole background, one would analyze the behavior of quantum fields within the wormhole's geometry. This requires solving the relevant covariant wave equations within the curved spacetime of the wormhole \cite{a1,a2,a3}. Similarly, in the context of black holes, effective dielectric properties can be inferred from how fields interact with the black hole's gravitational field. This analysis involves examining permittivity and permeability in the black hole's spacetime background. In each cases, determining the "effective dielectric properties" requires solving the field equations in the respective background and studying how fields interact with the material or spacetime geometry. From this viewpoint, it seems possible to assess the "effective dielectric properties" of a black hole's near-horizon background by employing methods like those documented in \cite{kobra}.

\section*{\small{Funding}}
This research has not received any funding.



\end{document}